\documentclass[reprint,superscriptaddress,amsmath,amsfonts,amssymb,aps,prl,showkeys]{revtex4-2}

\usepackage{hyperref}
\usepackage{graphicx}
\hypersetup{
  breaklinks = {true},
  citecolor = {blue},
  colorlinks = {true},
  linkcolor = {red},
}

\usepackage{soul}
\usepackage{physics}
\usepackage{mathtools}
\usepackage[dvipsnames]{xcolor}
\usepackage{braket} 
\usepackage{upgreek}

\usepackage{ulem}

\begin{document}

\title{Resilient Intraparticle Entanglement and its Manifestation in \\ Spin Dynamics of Disordered Dirac Materials}

\author{Jorge Martínez Romeral}
\affiliation{Catalan Institute of Nanoscience and Nanotechnology (ICN2), CSIC and BIST, Campus UAB, Bellaterra, 08193 Barcelona, Spain}
\affiliation{Department of Physics, Universitat Autònoma de Barcelona (UAB), Campus UAB, Bellaterra, 08193 Barcelona, Spain}
\author{Aron W. Cummings}
\affiliation{Catalan Institute of Nanoscience and Nanotechnology (ICN2), CSIC and BIST, Campus UAB, Bellaterra, 08193 Barcelona, Spain}
\author{Stephan Roche}
\affiliation{Catalan Institute of Nanoscience and Nanotechnology (ICN2), CSIC and BIST,
Campus UAB, Bellaterra, 08193 Barcelona, Spain}
\affiliation{ICREA--Instituci\'o Catalana de Recerca i Estudis Avan\c{c}ats, 08010 Barcelona, Spain}

\date{\today}

\begin{abstract}
Topological quantum matter exhibits novel transport phenomena driven by entanglement between internal degrees of freedom, as for instance generated by spin-orbit coupling effects. Here we report on a direct connection between the mechanism driving spin relaxation and the intertwined dynamics between spin and sublattice degrees of freedom in disordered graphene. Beyond having a direct observable consequence, such intraparticle entanglement is shown to be resilient to disorder, pointing towards a novel resource for quantum information processing.
\end{abstract}

\maketitle

The study of quantum transport in (topological) quantum matter is a central topic of modern condensed matter, given its connection to an emerging class of nontrivial phenomena and materials including topologically protected edge states, exotic Moir\'{e} physics, twisted multilayers of 2D materials, and strongly correlated systems \cite{Giustino_2020,Weber_2024}. Particularly relevant is the existence of the variety of internal degrees of freedom which arise from lattice symmetry and internal degeneracies, such as A/B sublattice pseudospin or $\mathbf{K}_+ / \mathbf{K}_-$ valley isospin in graphene, or layer pseudospin in multilayer systems assembled by stacking (and twist angle) engineering. The contributions of spin-orbit coupling, magnetic exchange fields and Coulomb interactions further yield nontrivial modifications of intraparticle and interparticle entanglement properties as well as the formation of strongly correlated and topological states, which result in a wealth of puzzling and exotic phenomena including superconductivity and orbital magnetism \cite{Andrei2021, Tian2023}.

In this wide context, exploration of the dynamics of internal degrees of freedom and their intertwined evolution is particularly relevant when a connection can be made with some directly accessible experimental observable. The case of graphene has been particularly scrutinized given the longstanding debate concerning the nature of the spin transport and spin relaxation mechanisms at play \cite{DPgraphene, PhysRevLett.103.146801, OchoaEygraphene, Zhang_2012, PhysRevLett.104.187201, PhysRevLett.107.047207, PhysRevLett.109.186604, PhysRevB.86.161416, PhysRevLett.112.116602, Soriano_2015, Drögeler2016, PhysRevB.92.195408,Tuan2014, VanTuan2016, PhysRevLett.116.086602, Gebeyehu_2019}, while entanglement effects on weak localization \cite{Sousa2022} or spin-orbit torque phenomena \cite{duenas2023emerging} have also been discussed recently. Prior literature has established that depending on the nature of disorder, spin relaxation may be dominated by magnetic impurities, the Elliott-Yafet mechanism, or the D'yakonov-Perel' mechanism, while the absolute values of spin lifetime mainly depend on the degree of disorder and the strength of the spin-orbit coupling induced in graphene by the substrate. Additionally, the minimum in spin lifetime around the charge neutrality point has been attributed to either resonant magnetic impurities or the EY mechanism. \cite{DPgraphene, PhysRevLett.103.146801,OchoaEygraphene,Zhang_2012,PhysRevLett.104.187201, PhysRevLett.107.047207, PhysRevLett.109.186604, PhysRevB.86.161416, PhysRevLett.112.116602, Soriano_2015, Drögeler2016, PhysRevB.92.195408,Tuan2014, VanTuan2016, PhysRevLett.116.086602, Gebeyehu_2019}.

The peculiar dynamics of intraparticle entanglement between spin and sublattice in massless Dirac fermions, generated by spin-orbit coupling, have also been argued to explain the energy dependence of spin lifetime in graphene around the charge neutrality point \cite{Tuan2014}. Although its utility in quantum information processing and quantum communication has not yet been clarified, given its magnitude and resilience to state preparation in the ultraclean limit \cite{PhysRevB.102.041403}, as well as proposals to transfer intraparticle entanglement to two-body interparticle entanglement \cite{Yonac2007, swapintrainter}, such spin-sublattice entanglement may offer a novel resource of quantum information by complementing other possible entanglement mechanisms \cite{PhysRevB.79.115444, PhysRevB.95.195145}. However, to date the dynamics and robustness of such intraparticle entanglement in the presence of disorder remains to be explored. Additionally, a direct connection between entanglement and the spin lifetime in graphene has yet to be elucidated.

Here, we use numerical simulations and analytical derivations to study the dynamics of spin and intraparticle entanglement in disordered graphene. We consider two types of disorder – charge impurities and magnetic impurities – and in both cases we find that the entanglement saturates to a finite value, independent of the initial state and the type or strength of disorder. This spin-sublattice entanglement thus appears to be an equilibrium property of graphene, with a magnitude that is robust to charge and spin scattering. For future proposals to use such intraparticle entanglement as a quantum resource, these results suggest that disorder may not be an inherently limiting factor.

Additionally, we derive an expression that directly relates the intraparticle entanglement to the Elliott-Yafet mechanism of spin relaxation in graphene. In the absence of magnetic impurities, this mechanism is dominant at low doping, yielding a minimum in the spin lifetime at the charge neutrality point. Our results therefore offer a direct, experimentally observable consequence of the innate spin-sublattice entanglement in graphene, and suggest that this connection could be used to monitor quantum information, as well as be generalized to other entanglement phenomena in similar materials.

{\it Electronic model and measure of entanglement ---}
To study its entanglement properties, we consider a continuum model of graphene near the $\mathbf{K}_{\pm}$ points with Rashba spin-orbit coupling (SOC) induced by a perpendicular electric field or a substrate \cite{PhysRevB.79.161409,Sierra2021}. The Hamiltonian is
\begin{align}\label{eq:Heff}
  \hat{H}(\mathbf{k}) &= \hbar v_{\mathrm{F}}\left(\nu\hat{\sigma}_xk_x+\hat{\sigma}_yk_y\right)\otimes \hat{s}_0 \nonumber \\
  &+ \lambda_{\mathrm{R}}\left(\nu\hat{\sigma}_x\otimes\hat{s}_y-\hat{\sigma}_y\otimes\hat{s}_x\right),
\end{align}
where $v_{\mathrm{F}}$ is the Fermi velocity, $\nu=\pm 1$ is the valley index, $\mathbf{k}$ is the crystalline momentum with respect to the $\mathbf{K}_{\pm}$ points, $\lambda_{\mathrm{R}}$ is the Rashba SOC strength and $\hat{\mathrm{s}}_i$ ($\hat{\mathrm{\sigma}}_i$) are the Pauli matrices in the spin (sublattice) space. The considered parameters ($E_\mathrm{F} \le 15$ meV and $\lambda_\mathrm{R} \le 500$ $\upmu$eV) fall within the range of the validity of the continuum model of graphene, which deviates from the tight-binding (TB) model for $E_\mathrm{F} \approx 300$ meV and large Rashba SOC \cite{CastroNeto2009, Zarea2009}. Simulations using the TB model leave our results unchanged. Without loss of generality and unless otherwise specified, we restrict ourselves to a single valley, $\nu=+1$. 

\begin{figure}[tbh]
    \centering
    \includegraphics[width=1\textwidth]{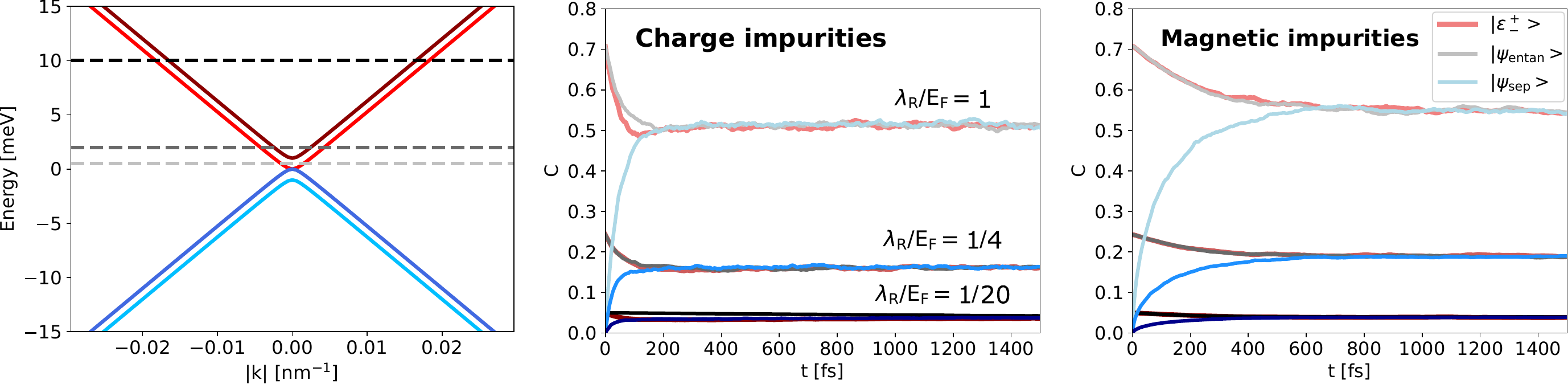}
    \caption{Left panel: graphene band structure for $\lambda_\mathrm{R}=500$ $\upmu$eV. The dashed lines indicate, from lighter to darker, the Fermi energies $E_{\mathrm{F}}=0.5$, $2$, and $10$ meV. Center panel: ensemble concurrence as a function of time for the three Fermi energies indicated in the left panel, for charge scattering time $\tau=50$ fs.
    The red lines correspond to the eigenstate of the lower conduction band $\ket{\varepsilon_-^+}$, the black/grey lines are for the state $\ket{\psi_{\mathrm{entan}}}$, and the blue lines are for  $\ket{\psi_{\mathrm{sep}}}$. Right panel: same as the center panel, but for magnetic impurity scattering.}
    \label{fig:newfig1}
\end{figure}

The eigenenergies of the system are {\small $\varepsilon_\pm^\xi(\mathbf{k}) = \xi \left( \pm\lambda_{\mathrm{R}} + \sqrt{\lambda_{\mathrm{R}}^2+\varepsilon_0^2}\right)$}, where $\xi=+ 1\,(-1)$ for electrons (holes) and $\varepsilon_0=\hbar v_{\mathrm{F}}|\mathbf{k}|$ is the energy dispersion in the absence of SOC. The band structure of this system, with $\lambda_{\mathrm{R}}=500$ $\upmu$eV, is shown in the left panel of Fig.\ \ref{fig:newfig1}. The eigenstates of the Hamiltonian, in the sublattice-spin basis $\left\{ A_\uparrow, A_\downarrow, B_\uparrow, B_\downarrow \right\}$, are given by
\begin{equation}\label{eq:eigenstates}
    \ket{\varepsilon^\xi_\pm (\mathbf{k})} = \dfrac{1}{\sqrt{N_\pm}} \begin{bmatrix} e^{-i\theta} & \pm i \gamma_\pm & \xi\gamma_\pm & \pm\xi i e^{i\theta} \end{bmatrix}^{\mathrm{T}},
\end{equation}
where $\theta = \atan(k_y/k_x)$ is the direction of momentum in the graphene plane, $\gamma_\pm = \varepsilon_\pm/\varepsilon_0$, and $N_\pm = 2(1+\gamma_\pm^2)$.

The entanglement between the spin sublattice of an arbitrary state $\ket{\psi}$ in the above basis can be quantified using the concurrence $C_{\psi}$ \cite{Wootters1998}. For pure states this is computed as
 \begin{equation}\label{eq:concurrence}
     C_{\psi} = | \braket{\psi|\widetilde{\psi}} | = | \bra{\psi} \left( \hat{\sigma}_Y \otimes \hat{\sigma}_Y \right) \ket{\psi^*} |,
 \end{equation}
where $\ket{\widetilde{\psi}} = \left( \hat{\sigma}_Y \otimes \hat{\sigma}_Y \right) \ket{\psi^*}$ denotes the spin-flip transformation of $\ket{\psi}$ in the Bloch sphere, $\ket{\psi^*}$ is the element-wise complex conjugate of $\ket{\psi}$, and $\hat{\sigma}_Y = \begin{bsmallmatrix} 0 & -i \\ i & 0 \end{bsmallmatrix}$ in spin space. For pure states of the form $\ket{\psi} = \begin{bmatrix} a&b&c&d \end{bmatrix}^{\mathrm{T}}$ that we consider here, $\ket{\psi^*} = \begin{bmatrix} a^*&b^*&c^*&d^* \end{bmatrix}^{\mathrm{T}}$, $\ket{\widetilde{\psi}} = \begin{bmatrix} -d^*&c^*&b^*&-a^* \end{bmatrix}^{\mathrm{T}}$, and thus $C_\psi = 2\left| ad-bc \right|$.
The concurrence is a so-called entanglement monotone, equal to 0 for separable states and 1 for maximally entangled states. Applying its definition to Eq.\ \eqref{eq:eigenstates}, we find that the concurrence of the eigenstates is $C_\varepsilon = \lambda_{\mathrm{R}} / \sqrt{\varepsilon_0^2+\lambda_{\mathrm{R}}^2}$. This entanglement emerges as the SOC effectively ``mixes" the spin and sublattice degrees of freedom, making their mean value disappear when the entanglement becomes stronger \cite{PhysRevB.102.041403}.

\begin{figure}[tbh]
\includegraphics[width=0.95\columnwidth]{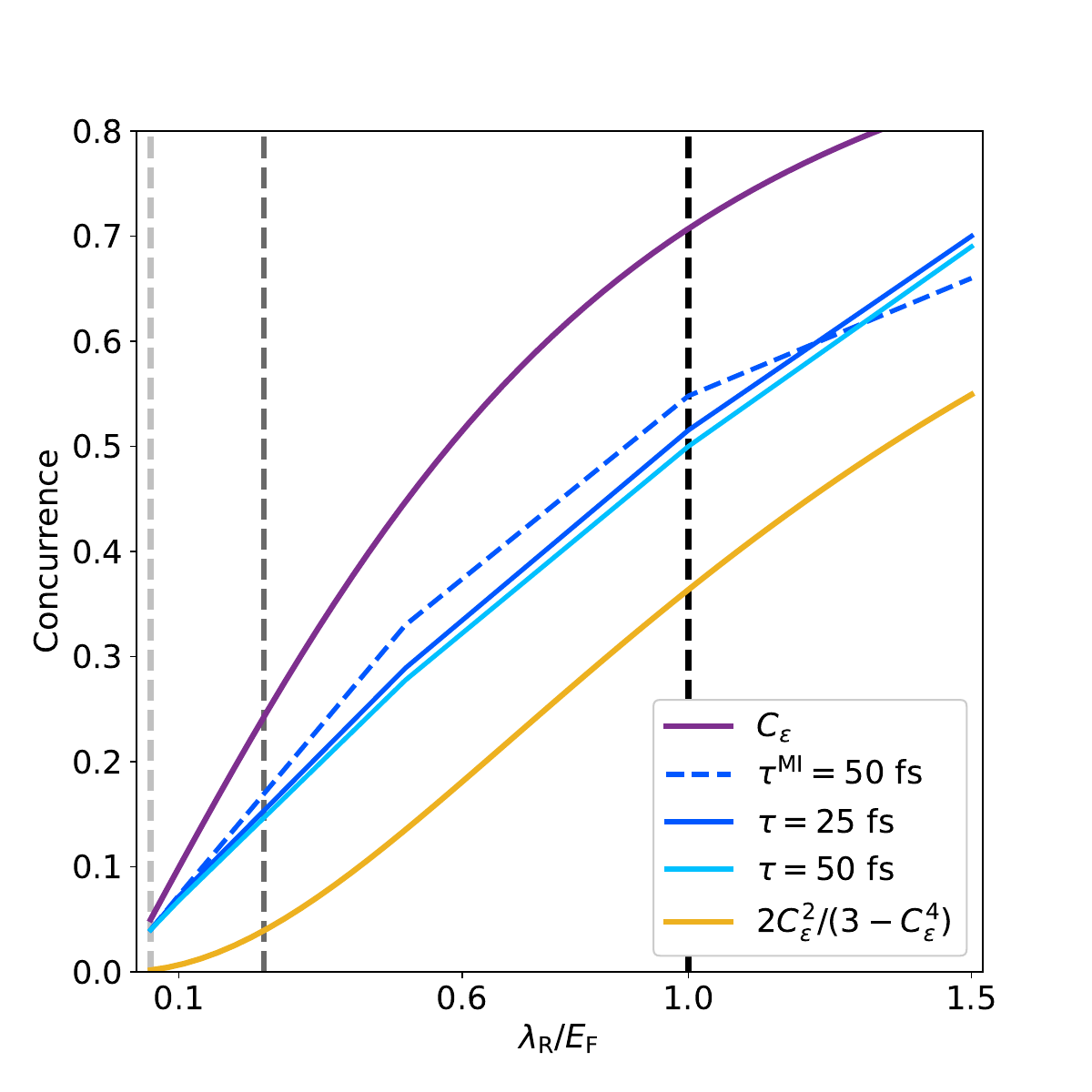}
\caption{\label{fig:figure1} Long-time concurrence for different scattering times in the presence of intravalley charge impurities (solid blue lines) and magnetic impurities (dashed blue line). The solid purple line indicates the concurrence of the eigenstates of the system and the upper bound of the converged concurrence, and the solid yellow line represents the lower bound of the converged concurrence. The vertical gray/black lines correspond to the Fermi energies of Fig.\ \ref{fig:newfig1}.}
\end{figure}

Therefore, the eigenstates of graphene in the presence of Rashba SOC are maximally entangled at the charge neutrality point ($\varepsilon_0 = 0$) and become separable for higher doping, with $C_\varepsilon$ decaying as $\sim$$1/\varepsilon_0$. On the other hand, increasing the Rashba coupling will increase the entanglement for a given Fermi energy ($E_{\mathrm{F}}$). In general, the concurrence is a monotonically-increasing function of $x = \lambda_\mathrm{R} / E_\mathrm{F}$, given by $C_\varepsilon = x / \sqrt{1+x^2}$, where we have let $E_\mathrm{F} \approx \varepsilon_0$. The purple solid line in Fig.\ \ref{fig:figure1} shows precisely this behavior.
\onecolumngrid

\twocolumngrid

{\it Entanglement dynamics in the presence of disorder ---}
In a previous work, we examined the dynamics of intraparticle spin-sublattice entanglement in perfectly clean graphene \cite{PhysRevB.102.041403}. Here we examine the dynamics and robustness of such intraparticle entanglement in the presence of disorder.
To do so, we have implemented a single-particle Monte Carlo method that tracks the concurrence and spin of a quantum state as it undergoes free flight and scattering in the presence of different types of disorder.
Our approach follows the general procedure for semiclassical Monte Carlo simulation of semiconductors \cite{Jacoboni1983} (details in the Supplemental Material \cite{suppmaterial}).
We consider the time evolution of a state at a given $E_{\mathrm{F}}$ (and thus at corresponding value of $\mathbf{k}$) either in the presence of charge impurities which randomize the momentum direction at every scattering event, or magnetic impurities which randomize the spin orientation. The impurity density and scattering strength are related to the scattering time, $\tau$, which is a free parameter in our simulations. Finally, we consider a large number of trajectories to calculate the ensemble average of the concurrence, $\braket{C(t)} = \frac{1}{N}\sum_{i=1}^N C_{\psi_i(t)}$, and the out-of-plane spin polarization, $\braket{s_z(t)} = \frac{1}{N}\sum_{i=1}^N\bra{\psi_i(t)}\hat{s}_z\ket{\psi_i(t)}$, as they evolve in time, where $N$ is the total number of particle trajectories, and $\ket{\psi_i(t)}$ is the state of the $i$th particle.

Some examples of the ensemble concurrence dynamics are shown in the center and right panels of Fig.\ \ref{fig:newfig1}, for $\tau=50$ fs and $\lambda_{\mathrm{R}}=500$ $\upmu$eV, at the Fermi energies indicated in the left panel. The center panel shows the dynamics arising from charge scattering, while in the right panel the dynamics are driven by magnetic impurities. The grey/black curves correspond to an initially fully-entangled Bell-type state, $\ket{\psi_\mathrm{entan}}=\frac{1}{\sqrt{2}}\left[1\,0\,0\,\mathrm{i}\right]^{\mathrm{T}}=\frac{1}{\sqrt{2}}\left(\ket{A}\otimes\ket{\uparrow}+\mathrm{i}\ket{B}\otimes\ket{\downarrow}\right)$, which is an eigenstate of the Hamiltonian at $|\mathbf{k}| = 0$. The blue curves correspond to an initially fully separable state $\ket{\psi_\mathrm{sep}}=\frac{1}{\sqrt{2}}\left[1\,0\,1\,0\right]^{\mathrm{T}}=\frac{1}{\sqrt{2}}\left(\ket{A}+\ket{B}\right)\otimes\ket{\uparrow}$, with out-of-plane spin and in-plane pseudospin pointing in the $x$ direction, parallel to the initial direction of transport. Such a state can be obtained by injecting charge carriers from a $z$-polarized ferromagnetic contact. The red lines correspond to the eigenstate of the lower conduction band, $\ket{\varepsilon_-^+}$, see Eq.\ \eqref{eq:eigenstates}.

In all cases the ensemble concurrence saturates to a finite value in the long time limit. This saturated value is independent of the initial state and the type of scattering, and depends only on the Fermi energy. This is also true for the lowest pair of curves, which have not yet fully saturated over the displayed time range.

In  Fig.\ \ref{fig:figure1}, we show the converged value of the concurrence as a function of $x = \lambda_\mathrm{R} / E_\mathrm{F}$. The solid blue lines correspond to charge impurity scattering with $\tau = 25$ and $50$ fs. Meanwhile, the blue dashed line shows the case for magnetic impurities with $\tau = 50$ fs. In addition, the situation where charge impurities induce strong intervalley scattering is also found to yield similar results (see Supplemental Material for details \cite{suppmaterial}). Notably, all of these curves fall on top of each other, indicating that the intraparticle entanglement appears to be robust and universal in the presence of disorder in graphene.

Fig.\ \ref{fig:figure1} shows that the converged value of concurrence grows monotonically with $\lambda_\mathrm{R} / E_\mathrm{F}$. This scaling behaviour follows that of the eigenstates themselves, whose concurrence is shown by the solid purple line. More concretely, the converged concurrence arises from a linear combination of both bands at $E_\mathrm{F}$ and is bounded in the range $\frac{2C_{\varepsilon}^2}{3-C_{\varepsilon}^4} \leq C \leq C_{\varepsilon}$, represented by the solid yellow and purple lines respectively (see the Supplemental Material for details of the analytical derivation \cite{suppmaterial}). These bounds show that the converged concurrence will always be nonzero except for $\lambda_{\mathrm{R}} / E_{\mathrm{F}} = 0$ when the eigenstates are all separable. We note that these bounds are obtained via the long-time average of a single state of our system, while the results obtained from the Monte Carlo procedure are obtained via an ensemble average. These also coincide with the concurrence derived from the full density matrix of the system. The agreement of these procedures suggests that the concurrence of this system of noninteracting particles obeys ergodicity.

Because both bounds only depend on the concurrence of the eigenstates, we can say that the converged value of the concurrence is a universal value that is a consequence of, and can be directly related to, the intraparticle entanglement of the eigenstates induced by the Rashba SOC. Additionally, the fact that these bounds depend only on the properties of the eigenstates -- and not on $\tau$, the type of scattering, or the initial state -- directly indicates the robustness of this effect to disorder and scattering.
The relaxation of the initial concurrence may be linked to the dynamics of the spin and the sublattice degrees of freedom, both of which relax to equilibrium values at long times \cite{Reimann2008, Short2011, Richter2019}.
These results thus demonstrate a general picture of entanglement dynamics, with the initial nonequilibrium entanglement relaxing to an equilibrium value determined solely by the concurrence of the eigenstates.

\begin{figure}[tbh]
\includegraphics[width=0.8\columnwidth]{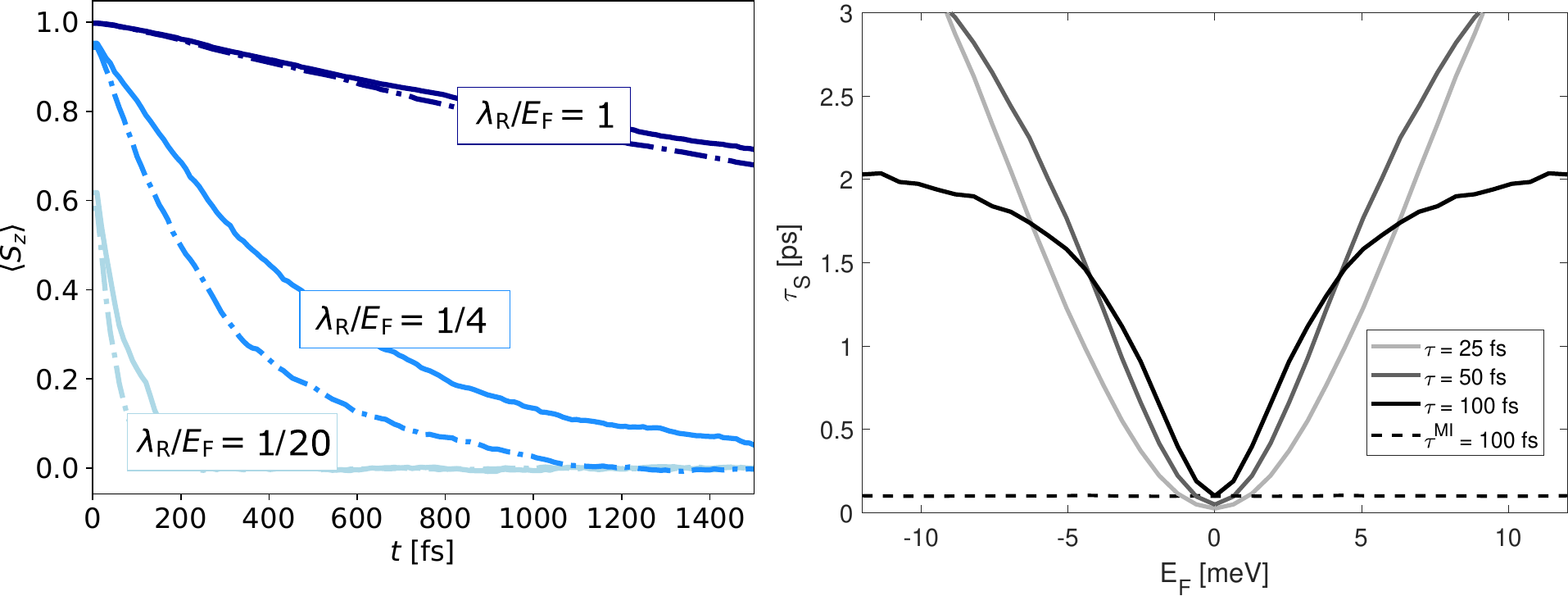}
\caption{\label{fig:figure2}Left panel: dynamics of the out-of-plane spin polarization of $\ket{\psi_\mathrm{sep}}$ in the presence of charge impurity scattering, for the three different Fermi energies of Fig.\ \ref{fig:newfig1}. The dot-dashed lines correspond to $\tau=25$ fs and the solid lines to $\tau=50$ fs. Right panel: spin relaxation time as a function of Fermi energy for different scattering times. Solid lines indicate charge impurity scattering and the dashed line indicates spin randomization by magnetic impurities.}
\end{figure}

{\it Spin relaxation time ---} Next we examine the features of the spin relaxation time of out-of-plane spins and directly relate it to the concurrence of the graphene eigenstates. The spin relaxation time, $\tau_\mathrm{s}$, is obtained from the Monte Carlo simulations by fitting the ensemble value of the out-of-plane spin polarization to either an exponential decay, $\braket{s_z(t)}=\exp(-t/\tau_\mathrm{s})$, or an oscillating exponential, $\braket{s_z(t)}=\exp(-t/\tau_\mathrm{s})\cos(2\pi t/T_{\Omega})$. We use the former when the spin precession time, $T_{\Omega} = \pi\hbar / \lambda_{\mathrm{R}}$, is longer than the scattering time and the latter when $T_{\Omega} < \tau$. The left panel of Fig.\ \ref{fig:figure2} shows examples of spin dynamics from the Monte Carlo simulations when $T_{\Omega} \gg \tau$, demonstrating the predicted exponential decay. Here we used $\ket{\psi_\mathrm{sep}}$ as the initial state, as it has a finite out-of-plane spin polarization. Meanwhile, the eigenstates of the system have zero out-of-plane spin polarization, and thus $\braket{s_z(t)}$ always relaxes to zero in the long-time limit.

In the right panel of Fig.\ \ref{fig:figure2}, we plot the spin relaxation time as a function of the Fermi energy for different values of the scattering time. As in Fig.\ \ref{fig:figure1}, the solid curves are for charge impurity scattering and the dashed curve is for magnetic impurities. The scattering times are given in the legend, and the Rashba SOC strength is $\lambda_\mathrm{R} = 500$ $\upmu$eV. The spin precession time is then $T_\Omega = \pi\hbar/\lambda_\mathrm{R} = 4.1$ ps, much longer than all values of $\tau$ that we consider.

For charge impurity scattering, at low energies $\tau_\mathrm{s}$ scales quadratically with $E_\mathrm{F}$ and is proportional to $\tau$. This is consistent with the Elliott-Yafet (EY) mechanism of spin relaxation, which in graphene was predicted to scale as $\tau_\mathrm{s}^\mathrm{EY} \propto (E_{\mathrm{F}} / \lambda_{\mathrm{R}})^2 \tau$ \cite{OchoaEygraphene}. At higher energies, when $\tau_\mathrm{s}^\mathrm{EY}$ becomes large, the spin dynamics are dominated by the D'yakonov-Perel (DP) mechanism, and the spin lifetime scales inversely with the scattering time, $\tau_\mathrm{s}^\mathrm{DP} = (\hbar/2\lambda_\mathrm{R})^2 / \tau$ \cite{DPgraphene}. This crossover between the EY and the DP regime happens at $E_\mathrm{F} \approx \hbar / \tau$.

In the case of scattering by magnetic impurities, $\tau_\mathrm{s}=\tau$ for all energies, as shown by the dashed line. This arises from the fact that magnetic impurities directly randomize the spin at each scattering event.

\onecolumngrid

\twocolumngrid
As seen in the right panel of Fig.\ \ref{fig:figure2}, at low energies the spin lifetime is dominated by EY relaxation, where $\tau_s \propto (E_\mathrm{F} / \lambda_\mathrm{R})^2$. Meanwhile, above we noted that the concurrence of the eigenstates scales approximately inversely with energy, $C_\varepsilon \approx (E_\mathrm{F} / \lambda_\mathrm{R})^{-1}$. In the following we demonstrate that there is a direct link between spin-sublattice entanglement and EY spin relaxation in graphene.

To derive an expression for EY spin relaxation in graphene, we start with the general interpretation of the mechanism -- that at every momentum scattering event, the spin polarization is reduced due to mixing of spin up and spin down states \cite{Zutic2004}. We define a spin loss coefficient that describes the relative amount of spin lost during the $n$th scattering event,
\begin{equation}\label{eq:coef}
\mathcal{S}^{n,n-1} \equiv \dfrac{\braket{s_z^{n}}}{\braket{s_z^{n-1}}},
\end{equation}
where $\braket{s_z^{n-1}}$ is the ensemble spin polarization just before the $n$th scattering event and $\braket{s_z^{n}}$ is the ensemble spin polarization just after. After a time $t$, the spin polarization will then be given by $\braket{s_z(t)} = \mathcal{S}^{N,N-1} \,\, ... \,\,\ \mathcal{S}^{2,1} \mathcal{S}^{1,0} \braket{s_z^0}$, where $t= N\tau$ with $\tau$ the average scattering interval. Assuming that every spin loss coefficient is equal, $\mathcal{S}^{n,n-1} \approx \mathcal{S}$, the net spin lost during an average interval $\tau$ is $\braket{s_z(t+\tau)} - \braket{s_z(t)} = -(1 - \mathcal{S}) \braket{s_z(t)}$, giving 
\begin{equation}
\frac{\mathrm{d}\braket{s_z(t)}}{\mathrm{d}t} \approx -\frac{1-\mathcal{S}}{\tau} \braket{s_z(t)}.
\end{equation}
This yields an exponential decay of the spin polarization, $\braket{s_z(t)} = \braket{s_z^0} \exp(-t/\tau_\mathrm{s}^\mathrm{EY})$, with spin lifetime
\begin{equation}\label{eq:ey_general}
\tau_\mathrm{s}^\mathrm{EY} = \frac{\tau}{1-\mathcal{S}}.
\end{equation}

To explicitly calculate $\mathcal{S}$, we start by considering the spin lost by a single particle during a single scattering event.
Before a scattering event, the state of a particle with momentum $\mathbf{k}$ can be written as a linear combination of the graphene+Rashba eigenstates, $\ket{\psi(\mathbf{k})} = \sum_{\xi\nu} \beta_{\xi\nu} \ket{\varepsilon_\nu^\xi (\mathbf{k})}$.
For $E_\mathrm{F} > 0$ $(E_\mathrm{F} < 0)$, this projection is over the conduction (valence) band eigenstates.
The spin polarization of this state is $\braket{s_z^\psi(\mathbf{k})} \equiv \braket{\psi(\mathbf{k}) | \hat{s}_z | \psi(\mathbf{k})}$.
A charge impurity scattering event changes $\mathbf{k} \rightarrow \mathbf{k'}$, and thus the scattered state can be written as a projection onto the eigenstates at $\mathbf{k'}$, $\ket{\psi(\mathbf{k'})} = \sum_{\xi\nu} \beta_{\xi\nu}' \ket{\varepsilon_\nu^\xi (\mathbf{k'})}$, where $\beta_{\xi\nu}' =  \braket{\varepsilon_\nu^\xi (\mathbf{k'}) | \psi(\mathbf{k})}$, with proper normalization.
The spin polarization of this scattered state is $\braket{s_z^\psi(\mathbf{k'})} = \braket{\psi(\mathbf{k'}) | \hat{s}_z | \psi(\mathbf{k'})}$, and the spin loss in this particular scattering event is thus $\mathcal{S}_\psi(\mathbf{k}-\mathbf{k'}) = \braket{s_z^\psi(\mathbf{k'})} / \braket{s_z^\psi(\mathbf{k})}$.
Finally, the average spin loss per scattering event, $\mathcal{S}$, is the average of $\mathcal{S}_\psi(\mathbf{k}-\mathbf{k'})$ over all $\mathbf{k}-\mathbf{k'}$, accounting for the form of $\ket{\psi}$ after a large number of scattering events.
By calculating $\mathcal{S}$ in this way, we then obtain the EY spin relaxation time by applying Eq.\ \eqref{eq:ey_general}.

Following this procedure (see Supplemental Material for details \cite{suppmaterial}), we arrive at the following expression for EY spin relaxation in graphene with Rashba SOC,
\begin{equation}\label{eq:ts_analytical}
    \tau_\mathrm{s}^\mathrm{EY} = \dfrac{3-C_\varepsilon^4}{5C_\varepsilon^2-3C_\varepsilon^4}\,\tau.
\end{equation}
This expression, with explicit dependence on the concurrence of the graphene eigenstates, can be broken down into two regimes. At high energies, when $C_\varepsilon \approx \lambda_\mathrm{R} / E_\mathrm{F} \ll 1$, Eq.\ \eqref{eq:ts_analytical} reduces to $\tau_\mathrm{s}^\mathrm{EY} \approx (3/5) (1 / C_\varepsilon)^2 \tau \approx (3/5) (E_\mathrm{F} / \lambda_\mathrm{R})^2 \tau$. This is the usual EY relation in graphene \cite{OchoaEygraphene}. On the other hand, for low energies when $E_{\mathrm{F}} \lesssim \lambda_{\mathrm{R}}$ and thus $C_\varepsilon \approx 1$, the EY relation becomes $\tau_\mathrm{s}^\mathrm{EY} \approx \tau$. In this regime, when there is only one band at $E_\mathrm{F}$, every scattering event changes the spin.

\begin{figure}[tbh]
\includegraphics[width=\columnwidth]{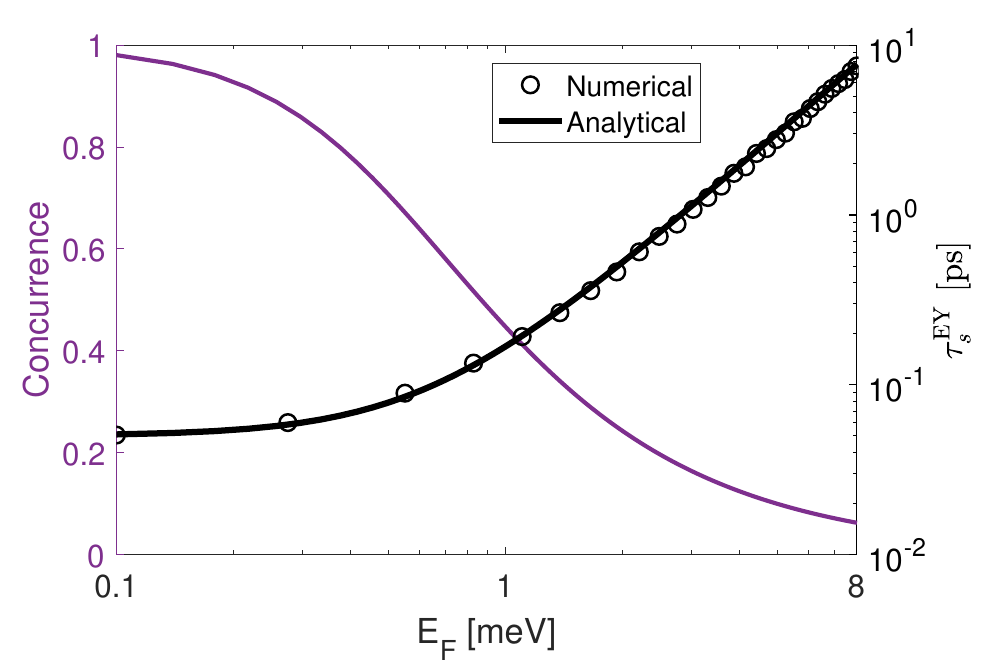}
\caption{\label{fig:figure3} Numerical and analytical spin relaxation times due to the EY mechanism (black, right axis), along with the concurrence (purple, left axis), as a function of the Fermi energy. The numerical results are from the Monte Carlo simulations, while the analytical spin lifetime is given by Eq.\ \eqref{eq:ts_analytical}.}
\end{figure}

The full scaling of $\tau_\mathrm{s}^\mathrm{EY}$ with Fermi energy is shown in Fig.\ \ref{fig:figure3} (right axis). The black circles are EY spin relaxation times extracted from the Monte Carlo simulations, while the solid black line corresponds to Eq.\ \eqref{eq:ts_analytical}, showing perfect agreement. On the left axis, we show the concurrence of the eigenstates, highlighting the inverse correlation between the spin-sublattice entanglement and EY spin relaxation in graphene.

{\it Conclusions ---} Using Monte Carlo simulations and analytical derivations, we have revealed the nature of spin-sublattice entanglement dynamics in disordered graphene. We find that this intraparticle entanglement evolves to a \textit{universal value} that is determined solely by the entanglement of the eigenstates of graphene induced by Rashba SOC. This universal value is independent of the initial state, is independent of the type of disorder (charge impurity scattering vs. spin randomization by magnetic impurities), and crucially, is independent of the disorder strength via the scattering time $\tau$. Being independent of $\tau$, the converged concurrence will thus be independent of transport quantities such as conductance or mobility. As indicated in Figs.\ \ref{fig:newfig1} and \ref{fig:figure1}, it depends only on the ratio $\lambda_\mathrm{R}/E_\mathrm{F}$, and thus is an intrinsic property of graphene with Rashba spin-orbit coupling. These results suggest that potential applications of this intraparticle entanglement as a resource may not be inherently limited by disorder. However, one could anticipate that spatial fluctuations in the Rashba SOC might reduce the magnitude of this converged value of entanglement.

Next, we have derived an explicit relation between Eliott-Yafet spin relaxation and the spin-sublattice entanglement in graphene. When spin relaxation is driven by spin-orbit coupling, the EY mechanism is dominant at low doping, yielding a minimum of spin lifetime at the charge neutrality point. Thus, the behavior of spin transport at low doping in graphene may be a direct experimental signature of the quantum entanglement properties of the charge carriers.

Beyond the quantification of entanglement through experimental data, one may also envision novel schemes of data processing and storage that utilize the multiple internal degrees of freedom of quantum matter. While the use of intraparticle entanglement as a quantum resource is not yet known, some work has suggested that it may be swapped onto two-body interparticle entanglement \cite{Yonac2007,swapintrainter}, which is a known quantum resource. Looking ahead, the complex interplay between intraparticle and interparticle entanglement remains to be further explored, with this work laying the groundwork for understanding such dynamics in graphene.

Beyond spin-sublattice intraparticle entanglement in single-layer graphene, we expect similar features to arise in other Dirac systems that carry multiple internal degrees of freedom, such as sublattice-layer entanglement in bilayer graphene \cite{PhysRevB.95.195145,Nguyen_2022}, spin-sublattice-layer entanglement in multilayer graphene structures with spin-orbit coupling, or spin-valley entanglement in layered transition metal dichalcogenides.

\begin{acknowledgments}
ICN2 is funded by the CERCA programme / Generalitat de Catalunya, and is supported by the Severo Ochoa Centres of Excellence programme, Grant CEX2021-001214-S, funded by MCIN / AEI / 10.13039.501100011033. This work is also supported by MICIN with European funds -- NextGenerationEU (PRTR‐C17.I1) and by Generalitat de Catalunya.
\end{acknowledgments}


\begin{thebibliography}{39}%
\makeatletter
\providecommand \@ifxundefined [1]{%
 \@ifx{#1\undefined}
}%
\providecommand \@ifnum [1]{%
 \ifnum #1\expandafter \@firstoftwo
 \else \expandafter \@secondoftwo
 \fi
}%
\providecommand \@ifx [1]{%
 \ifx #1\expandafter \@firstoftwo
 \else \expandafter \@secondoftwo
 \fi
}%
\providecommand \natexlab [1]{#1}%
\providecommand \enquote  [1]{``#1''}%
\providecommand \bibnamefont  [1]{#1}%
\providecommand \bibfnamefont [1]{#1}%
\providecommand \citenamefont [1]{#1}%
\providecommand \href@noop [0]{\@secondoftwo}%
\providecommand \href [0]{\begingroup \@sanitize@url \@href}%
\providecommand \@href[1]{\@@startlink{#1}\@@href}%
\providecommand \@@href[1]{\endgroup#1\@@endlink}%
\providecommand \@sanitize@url [0]{\catcode `\\12\catcode `\$12\catcode `\&12\catcode `\#12\catcode `\^12\catcode `\_12\catcode `\%12\relax}%
\providecommand \@@startlink[1]{}%
\providecommand \@@endlink[0]{}%
\providecommand \url  [0]{\begingroup\@sanitize@url \@url }%
\providecommand \@url [1]{\endgroup\@href {#1}{\urlprefix }}%
\providecommand \urlprefix  [0]{URL }%
\providecommand \Eprint [0]{\href }%
\providecommand \doibase [0]{https://doi.org/}%
\providecommand \selectlanguage [0]{\@gobble}%
\providecommand \bibinfo  [0]{\@secondoftwo}%
\providecommand \bibfield  [0]{\@secondoftwo}%
\providecommand \translation [1]{[#1]}%
\providecommand \BibitemOpen [0]{}%
\providecommand \bibitemStop [0]{}%
\providecommand \bibitemNoStop [0]{.\EOS\space}%
\providecommand \EOS [0]{\spacefactor3000\relax}%
\providecommand \BibitemShut  [1]{\csname bibitem#1\endcsname}%
\let\auto@bib@innerbib\@empty
\bibitem [{\citenamefont {Giustino}\ \emph {et~al.}(2021)\citenamefont {Giustino}, \citenamefont {Lee}, \citenamefont {Trier}, \citenamefont {Bibes}, \citenamefont {Winter}, \citenamefont {Valentí}, \citenamefont {Son}, \citenamefont {Taillefer}, \citenamefont {Heil}, \citenamefont {Figueroa}, \citenamefont {Plaçais}, \citenamefont {Wu}, \citenamefont {Yazyev}, \citenamefont {Bakkers}, \citenamefont {Nygård}, \citenamefont {Forn-Díaz}, \citenamefont {Franceschi}, \citenamefont {McIver}, \citenamefont {Torres}, \citenamefont {Low}, \citenamefont {Kumar}, \citenamefont {Galceran}, \citenamefont {Valenzuela}, \citenamefont {Costache}, \citenamefont {Manchon}, \citenamefont {Kim}, \citenamefont {Schleder}, \citenamefont {Fazzio},\ and\ \citenamefont {Roche}}]{Giustino_2020}%
  \BibitemOpen
  \bibfield  {author} {\bibinfo {author} {\bibfnamefont {F.}~\bibnamefont {Giustino}}, \bibinfo {author} {\bibfnamefont {J.~H.}\ \bibnamefont {Lee}}, \bibinfo {author} {\bibfnamefont {F.}~\bibnamefont {Trier}}, \bibinfo {author} {\bibfnamefont {M.}~\bibnamefont {Bibes}}, \bibinfo {author} {\bibfnamefont {S.~M.}\ \bibnamefont {Winter}}, \bibinfo {author} {\bibfnamefont {R.}~\bibnamefont {Valentí}}, \bibinfo {author} {\bibfnamefont {Y.-W.}\ \bibnamefont {Son}}, \bibinfo {author} {\bibfnamefont {L.}~\bibnamefont {Taillefer}}, \bibinfo {author} {\bibfnamefont {C.}~\bibnamefont {Heil}}, \bibinfo {author} {\bibfnamefont {A.~I.}\ \bibnamefont {Figueroa}}, \bibinfo {author} {\bibfnamefont {B.}~\bibnamefont {Plaçais}}, \bibinfo {author} {\bibfnamefont {Q.}~\bibnamefont {Wu}}, \bibinfo {author} {\bibfnamefont {O.~V.}\ \bibnamefont {Yazyev}}, \bibinfo {author} {\bibfnamefont {E.~P. A.~M.}\ \bibnamefont {Bakkers}}, \bibinfo {author} {\bibfnamefont {J.}~\bibnamefont {Nygård}}, \bibinfo {author} {\bibfnamefont
  {P.}~\bibnamefont {Forn-Díaz}}, \bibinfo {author} {\bibfnamefont {S.~D.}\ \bibnamefont {Franceschi}}, \bibinfo {author} {\bibfnamefont {J.~W.}\ \bibnamefont {McIver}}, \bibinfo {author} {\bibfnamefont {L.~E. F.~F.}\ \bibnamefont {Torres}}, \bibinfo {author} {\bibfnamefont {T.}~\bibnamefont {Low}}, \bibinfo {author} {\bibfnamefont {A.}~\bibnamefont {Kumar}}, \bibinfo {author} {\bibfnamefont {R.}~\bibnamefont {Galceran}}, \bibinfo {author} {\bibfnamefont {S.~O.}\ \bibnamefont {Valenzuela}}, \bibinfo {author} {\bibfnamefont {M.~V.}\ \bibnamefont {Costache}}, \bibinfo {author} {\bibfnamefont {A.}~\bibnamefont {Manchon}}, \bibinfo {author} {\bibfnamefont {E.-A.}\ \bibnamefont {Kim}}, \bibinfo {author} {\bibfnamefont {G.~R.}\ \bibnamefont {Schleder}}, \bibinfo {author} {\bibfnamefont {A.}~\bibnamefont {Fazzio}},\ and\ \bibinfo {author} {\bibfnamefont {S.}~\bibnamefont {Roche}},\ }\bibfield  {title} {\bibinfo {title} {The 2021 quantum materials roadmap},\ }\href {https://doi.org/10.1088/2515-7639/abb74e}
  {\bibfield  {journal} {\bibinfo  {journal} {J. Phys. Mater.}\ }\textbf {\bibinfo {volume} {3}},\ \bibinfo {pages} {042006} (\bibinfo {year} {2021})}\BibitemShut {NoStop}%
\bibitem [{\citenamefont {Weber}\ \emph {et~al.}(2024)\citenamefont {Weber}, \citenamefont {Fuhrer}, \citenamefont {Sheng}, \citenamefont {Yang}, \citenamefont {Thomale}, \citenamefont {Shamim}, \citenamefont {Molenkamp}, \citenamefont {Cobden}, \citenamefont {Pesin}, \citenamefont {Zandvliet}, \citenamefont {Bampoulis}, \citenamefont {Claessen}, \citenamefont {Menges}, \citenamefont {Gooth}, \citenamefont {Felser}, \citenamefont {Shekhar}, \citenamefont {Tadich}, \citenamefont {Zhao}, \citenamefont {Edmonds}, \citenamefont {Jia}, \citenamefont {Bieniek}, \citenamefont {Väyrynen}, \citenamefont {Culcer}, \citenamefont {Muralidharan},\ and\ \citenamefont {Nadeem}}]{Weber_2024}%
  \BibitemOpen
  \bibfield  {author} {\bibinfo {author} {\bibfnamefont {B.}~\bibnamefont {Weber}}, \bibinfo {author} {\bibfnamefont {M.~S.}\ \bibnamefont {Fuhrer}}, \bibinfo {author} {\bibfnamefont {X.-L.}\ \bibnamefont {Sheng}}, \bibinfo {author} {\bibfnamefont {S.~A.}\ \bibnamefont {Yang}}, \bibinfo {author} {\bibfnamefont {R.}~\bibnamefont {Thomale}}, \bibinfo {author} {\bibfnamefont {S.}~\bibnamefont {Shamim}}, \bibinfo {author} {\bibfnamefont {L.~W.}\ \bibnamefont {Molenkamp}}, \bibinfo {author} {\bibfnamefont {D.}~\bibnamefont {Cobden}}, \bibinfo {author} {\bibfnamefont {D.}~\bibnamefont {Pesin}}, \bibinfo {author} {\bibfnamefont {H.~J.~W.}\ \bibnamefont {Zandvliet}}, \bibinfo {author} {\bibfnamefont {P.}~\bibnamefont {Bampoulis}}, \bibinfo {author} {\bibfnamefont {R.}~\bibnamefont {Claessen}}, \bibinfo {author} {\bibfnamefont {F.~R.}\ \bibnamefont {Menges}}, \bibinfo {author} {\bibfnamefont {J.}~\bibnamefont {Gooth}}, \bibinfo {author} {\bibfnamefont {C.}~\bibnamefont {Felser}}, \bibinfo {author} {\bibfnamefont
  {C.}~\bibnamefont {Shekhar}}, \bibinfo {author} {\bibfnamefont {A.}~\bibnamefont {Tadich}}, \bibinfo {author} {\bibfnamefont {M.}~\bibnamefont {Zhao}}, \bibinfo {author} {\bibfnamefont {M.~T.}\ \bibnamefont {Edmonds}}, \bibinfo {author} {\bibfnamefont {J.}~\bibnamefont {Jia}}, \bibinfo {author} {\bibfnamefont {M.}~\bibnamefont {Bieniek}}, \bibinfo {author} {\bibfnamefont {J.~I.}\ \bibnamefont {Väyrynen}}, \bibinfo {author} {\bibfnamefont {D.}~\bibnamefont {Culcer}}, \bibinfo {author} {\bibfnamefont {B.}~\bibnamefont {Muralidharan}},\ and\ \bibinfo {author} {\bibfnamefont {M.}~\bibnamefont {Nadeem}},\ }\bibfield  {title} {\bibinfo {title} {2024 roadmap on 2d topological insulators},\ }\href {https://doi.org/10.1088/2515-7639/ad2083} {\bibfield  {journal} {\bibinfo  {journal} {J. Phys. Mater.}\ }\textbf {\bibinfo {volume} {7}},\ \bibinfo {pages} {022501} (\bibinfo {year} {2024})}\BibitemShut {NoStop}%
\bibitem [{\citenamefont {Andrei}\ \emph {et~al.}(2021)\citenamefont {Andrei}, \citenamefont {Efetov}, \citenamefont {Jarillo-Herrero}, \citenamefont {MacDonald}, \citenamefont {Mak}, \citenamefont {Senthil}, \citenamefont {Tutuc}, \citenamefont {Yazdani},\ and\ \citenamefont {Young}}]{Andrei2021}%
  \BibitemOpen
  \bibfield  {author} {\bibinfo {author} {\bibfnamefont {E.~Y.}\ \bibnamefont {Andrei}}, \bibinfo {author} {\bibfnamefont {D.~K.}\ \bibnamefont {Efetov}}, \bibinfo {author} {\bibfnamefont {P.}~\bibnamefont {Jarillo-Herrero}}, \bibinfo {author} {\bibfnamefont {A.~H.}\ \bibnamefont {MacDonald}}, \bibinfo {author} {\bibfnamefont {K.~F.}\ \bibnamefont {Mak}}, \bibinfo {author} {\bibfnamefont {T.}~\bibnamefont {Senthil}}, \bibinfo {author} {\bibfnamefont {E.}~\bibnamefont {Tutuc}}, \bibinfo {author} {\bibfnamefont {A.}~\bibnamefont {Yazdani}},\ and\ \bibinfo {author} {\bibfnamefont {A.~F.}\ \bibnamefont {Young}},\ }\bibfield  {title} {\bibinfo {title} {The marvels of moir{\'e} materials},\ }\href {https://doi.org/10.1038/s41578-021-00284-1} {\bibfield  {journal} {\bibinfo  {journal} {Nat. Rev. Mater.}\ }\textbf {\bibinfo {volume} {6}},\ \bibinfo {pages} {201} (\bibinfo {year} {2021})}\BibitemShut {NoStop}%
\bibitem [{\citenamefont {Tian}\ \emph {et~al.}(2023)\citenamefont {Tian}, \citenamefont {Gao}, \citenamefont {Zhang}, \citenamefont {Che}, \citenamefont {Xu}, \citenamefont {Cheung}, \citenamefont {Watanabe}, \citenamefont {Taniguchi}, \citenamefont {Randeria}, \citenamefont {Zhang}, \citenamefont {Lau},\ and\ \citenamefont {Bockrath}}]{Tian2023}%
  \BibitemOpen
  \bibfield  {author} {\bibinfo {author} {\bibfnamefont {H.}~\bibnamefont {Tian}}, \bibinfo {author} {\bibfnamefont {X.}~\bibnamefont {Gao}}, \bibinfo {author} {\bibfnamefont {Y.}~\bibnamefont {Zhang}}, \bibinfo {author} {\bibfnamefont {S.}~\bibnamefont {Che}}, \bibinfo {author} {\bibfnamefont {T.}~\bibnamefont {Xu}}, \bibinfo {author} {\bibfnamefont {P.}~\bibnamefont {Cheung}}, \bibinfo {author} {\bibfnamefont {K.}~\bibnamefont {Watanabe}}, \bibinfo {author} {\bibfnamefont {T.}~\bibnamefont {Taniguchi}}, \bibinfo {author} {\bibfnamefont {M.}~\bibnamefont {Randeria}}, \bibinfo {author} {\bibfnamefont {F.}~\bibnamefont {Zhang}}, \bibinfo {author} {\bibfnamefont {C.~N.}\ \bibnamefont {Lau}},\ and\ \bibinfo {author} {\bibfnamefont {M.~W.}\ \bibnamefont {Bockrath}},\ }\bibfield  {title} {\bibinfo {title} {Evidence for dirac flat band superconductivity enabled by quantum geometry},\ }\href {https://doi.org/10.1038/s41586-022-05576-2} {\bibfield  {journal} {\bibinfo  {journal} {Nature}\ }\textbf {\bibinfo {volume}
  {614}},\ \bibinfo {pages} {440} (\bibinfo {year} {2023})}\BibitemShut {NoStop}%
\bibitem [{\citenamefont {Ertler}\ \emph {et~al.}(2009)\citenamefont {Ertler}, \citenamefont {Konschuh}, \citenamefont {Gmitra},\ and\ \citenamefont {Fabian}}]{DPgraphene}%
  \BibitemOpen
  \bibfield  {author} {\bibinfo {author} {\bibfnamefont {C.}~\bibnamefont {Ertler}}, \bibinfo {author} {\bibfnamefont {S.}~\bibnamefont {Konschuh}}, \bibinfo {author} {\bibfnamefont {M.}~\bibnamefont {Gmitra}},\ and\ \bibinfo {author} {\bibfnamefont {J.}~\bibnamefont {Fabian}},\ }\bibfield  {title} {\bibinfo {title} {Electron spin relaxation in graphene: The role of the substrate},\ }\href {https://doi.org/10.1103/PhysRevB.80.041405} {\bibfield  {journal} {\bibinfo  {journal} {Phys. Rev. B}\ }\textbf {\bibinfo {volume} {80}},\ \bibinfo {pages} {041405} (\bibinfo {year} {2009})}\BibitemShut {NoStop}%
\bibitem [{\citenamefont {Huertas-Hernando}\ \emph {et~al.}(2009)\citenamefont {Huertas-Hernando}, \citenamefont {Guinea},\ and\ \citenamefont {Brataas}}]{PhysRevLett.103.146801}%
  \BibitemOpen
  \bibfield  {author} {\bibinfo {author} {\bibfnamefont {D.}~\bibnamefont {Huertas-Hernando}}, \bibinfo {author} {\bibfnamefont {F.}~\bibnamefont {Guinea}},\ and\ \bibinfo {author} {\bibfnamefont {A.}~\bibnamefont {Brataas}},\ }\bibfield  {title} {\bibinfo {title} {Spin-orbit-mediated spin relaxation in graphene},\ }\href {https://doi.org/10.1103/PhysRevLett.103.146801} {\bibfield  {journal} {\bibinfo  {journal} {Phys. Rev. Lett.}\ }\textbf {\bibinfo {volume} {103}},\ \bibinfo {pages} {146801} (\bibinfo {year} {2009})}\BibitemShut {NoStop}%
\bibitem [{\citenamefont {Ochoa}\ \emph {et~al.}(2012)\citenamefont {Ochoa}, \citenamefont {Castro~Neto},\ and\ \citenamefont {Guinea}}]{OchoaEygraphene}%
  \BibitemOpen
  \bibfield  {author} {\bibinfo {author} {\bibfnamefont {H.}~\bibnamefont {Ochoa}}, \bibinfo {author} {\bibfnamefont {A.~H.}\ \bibnamefont {Castro~Neto}},\ and\ \bibinfo {author} {\bibfnamefont {F.}~\bibnamefont {Guinea}},\ }\bibfield  {title} {\bibinfo {title} {Elliot-yafet mechanism in graphene},\ }\href {https://doi.org/10.1103/PhysRevLett.108.206808} {\bibfield  {journal} {\bibinfo  {journal} {Phys. Rev. Lett.}\ }\textbf {\bibinfo {volume} {108}},\ \bibinfo {pages} {206808} (\bibinfo {year} {2012})}\BibitemShut {NoStop}%
\bibitem [{\citenamefont {Zhang}\ and\ \citenamefont {Wu}(2012)}]{Zhang_2012}%
  \BibitemOpen
  \bibfield  {author} {\bibinfo {author} {\bibfnamefont {P.}~\bibnamefont {Zhang}}\ and\ \bibinfo {author} {\bibfnamefont {M.~W.}\ \bibnamefont {Wu}},\ }\bibfield  {title} {\bibinfo {title} {Electron spin relaxation in graphene with random rashba field: comparison of the d'yakonov–perel' and elliott–yafet-like mechanisms},\ }\href {https://doi.org/10.1088/1367-2630/14/3/033015} {\bibfield  {journal} {\bibinfo  {journal} {New J. Phys.}\ }\textbf {\bibinfo {volume} {14}},\ \bibinfo {pages} {033015} (\bibinfo {year} {2012})}\BibitemShut {NoStop}%
\bibitem [{\citenamefont {Pi}\ \emph {et~al.}(2010)\citenamefont {Pi}, \citenamefont {Han}, \citenamefont {McCreary}, \citenamefont {Swartz}, \citenamefont {Li},\ and\ \citenamefont {Kawakami}}]{PhysRevLett.104.187201}%
  \BibitemOpen
  \bibfield  {author} {\bibinfo {author} {\bibfnamefont {K.}~\bibnamefont {Pi}}, \bibinfo {author} {\bibfnamefont {W.}~\bibnamefont {Han}}, \bibinfo {author} {\bibfnamefont {K.~M.}\ \bibnamefont {McCreary}}, \bibinfo {author} {\bibfnamefont {A.~G.}\ \bibnamefont {Swartz}}, \bibinfo {author} {\bibfnamefont {Y.}~\bibnamefont {Li}},\ and\ \bibinfo {author} {\bibfnamefont {R.~K.}\ \bibnamefont {Kawakami}},\ }\bibfield  {title} {\bibinfo {title} {Manipulation of spin transport in graphene by surface chemical doping},\ }\href {https://doi.org/10.1103/PhysRevLett.104.187201} {\bibfield  {journal} {\bibinfo  {journal} {Phys. Rev. Lett.}\ }\textbf {\bibinfo {volume} {104}},\ \bibinfo {pages} {187201} (\bibinfo {year} {2010})}\BibitemShut {NoStop}%
\bibitem [{\citenamefont {Han}\ and\ \citenamefont {Kawakami}(2011)}]{PhysRevLett.107.047207}%
  \BibitemOpen
  \bibfield  {author} {\bibinfo {author} {\bibfnamefont {W.}~\bibnamefont {Han}}\ and\ \bibinfo {author} {\bibfnamefont {R.~K.}\ \bibnamefont {Kawakami}},\ }\bibfield  {title} {\bibinfo {title} {Spin relaxation in single-layer and bilayer graphene},\ }\href {https://doi.org/10.1103/PhysRevLett.107.047207} {\bibfield  {journal} {\bibinfo  {journal} {Phys. Rev. Lett.}\ }\textbf {\bibinfo {volume} {107}},\ \bibinfo {pages} {047207} (\bibinfo {year} {2011})}\BibitemShut {NoStop}%
\bibitem [{\citenamefont {McCreary}\ \emph {et~al.}(2012)\citenamefont {McCreary}, \citenamefont {Swartz}, \citenamefont {Han}, \citenamefont {Fabian},\ and\ \citenamefont {Kawakami}}]{PhysRevLett.109.186604}%
  \BibitemOpen
  \bibfield  {author} {\bibinfo {author} {\bibfnamefont {K.~M.}\ \bibnamefont {McCreary}}, \bibinfo {author} {\bibfnamefont {A.~G.}\ \bibnamefont {Swartz}}, \bibinfo {author} {\bibfnamefont {W.}~\bibnamefont {Han}}, \bibinfo {author} {\bibfnamefont {J.}~\bibnamefont {Fabian}},\ and\ \bibinfo {author} {\bibfnamefont {R.~K.}\ \bibnamefont {Kawakami}},\ }\bibfield  {title} {\bibinfo {title} {Magnetic moment formation in graphene detected by scattering of pure spin currents},\ }\href {https://doi.org/10.1103/PhysRevLett.109.186604} {\bibfield  {journal} {\bibinfo  {journal} {Phys. Rev. Lett.}\ }\textbf {\bibinfo {volume} {109}},\ \bibinfo {pages} {186604} (\bibinfo {year} {2012})}\BibitemShut {NoStop}%
\bibitem [{\citenamefont {Zomer}\ \emph {et~al.}(2012)\citenamefont {Zomer}, \citenamefont {Guimar\~aes}, \citenamefont {Tombros},\ and\ \citenamefont {van Wees}}]{PhysRevB.86.161416}%
  \BibitemOpen
  \bibfield  {author} {\bibinfo {author} {\bibfnamefont {P.~J.}\ \bibnamefont {Zomer}}, \bibinfo {author} {\bibfnamefont {M.~H.~D.}\ \bibnamefont {Guimar\~aes}}, \bibinfo {author} {\bibfnamefont {N.}~\bibnamefont {Tombros}},\ and\ \bibinfo {author} {\bibfnamefont {B.~J.}\ \bibnamefont {van Wees}},\ }\bibfield  {title} {\bibinfo {title} {Long-distance spin transport in high-mobility graphene on hexagonal boron nitride},\ }\href {https://doi.org/10.1103/PhysRevB.86.161416} {\bibfield  {journal} {\bibinfo  {journal} {Phys. Rev. B}\ }\textbf {\bibinfo {volume} {86}},\ \bibinfo {pages} {161416} (\bibinfo {year} {2012})}\BibitemShut {NoStop}%
\bibitem [{\citenamefont {Kochan}\ \emph {et~al.}(2014)\citenamefont {Kochan}, \citenamefont {Gmitra},\ and\ \citenamefont {Fabian}}]{PhysRevLett.112.116602}%
  \BibitemOpen
  \bibfield  {author} {\bibinfo {author} {\bibfnamefont {D.}~\bibnamefont {Kochan}}, \bibinfo {author} {\bibfnamefont {M.}~\bibnamefont {Gmitra}},\ and\ \bibinfo {author} {\bibfnamefont {J.}~\bibnamefont {Fabian}},\ }\bibfield  {title} {\bibinfo {title} {Spin relaxation mechanism in graphene: Resonant scattering by magnetic impurities},\ }\href {https://doi.org/10.1103/PhysRevLett.112.116602} {\bibfield  {journal} {\bibinfo  {journal} {Phys. Rev. Lett.}\ }\textbf {\bibinfo {volume} {112}},\ \bibinfo {pages} {116602} (\bibinfo {year} {2014})}\BibitemShut {NoStop}%
\bibitem [{\citenamefont {Soriano}\ \emph {et~al.}(2015)\citenamefont {Soriano}, \citenamefont {Tuan}, \citenamefont {Dubois}, \citenamefont {Gmitra}, \citenamefont {Cummings}, \citenamefont {Kochan}, \citenamefont {Ortmann}, \citenamefont {Charlier}, \citenamefont {Fabian},\ and\ \citenamefont {Roche}}]{Soriano_2015}%
  \BibitemOpen
  \bibfield  {author} {\bibinfo {author} {\bibfnamefont {D.}~\bibnamefont {Soriano}}, \bibinfo {author} {\bibfnamefont {D.~V.}\ \bibnamefont {Tuan}}, \bibinfo {author} {\bibfnamefont {S.~M.-M.}\ \bibnamefont {Dubois}}, \bibinfo {author} {\bibfnamefont {M.}~\bibnamefont {Gmitra}}, \bibinfo {author} {\bibfnamefont {A.~W.}\ \bibnamefont {Cummings}}, \bibinfo {author} {\bibfnamefont {D.}~\bibnamefont {Kochan}}, \bibinfo {author} {\bibfnamefont {F.}~\bibnamefont {Ortmann}}, \bibinfo {author} {\bibfnamefont {J.-C.}\ \bibnamefont {Charlier}}, \bibinfo {author} {\bibfnamefont {J.}~\bibnamefont {Fabian}},\ and\ \bibinfo {author} {\bibfnamefont {S.}~\bibnamefont {Roche}},\ }\bibfield  {title} {\bibinfo {title} {Spin transport in hydrogenated graphene},\ }\href {https://doi.org/10.1088/2053-1583/2/2/022002} {\bibfield  {journal} {\bibinfo  {journal} {2D Mater.}\ }\textbf {\bibinfo {volume} {2}},\ \bibinfo {pages} {022002} (\bibinfo {year} {2015})}\BibitemShut {NoStop}%
\bibitem [{\citenamefont {Dr{\"o}geler}\ \emph {et~al.}(2016)\citenamefont {Dr{\"o}geler}, \citenamefont {Franzen}, \citenamefont {Volmer}, \citenamefont {Pohlmann}, \citenamefont {Banszerus}, \citenamefont {Wolter}, \citenamefont {Watanabe}, \citenamefont {Taniguchi}, \citenamefont {Stampfer},\ and\ \citenamefont {Beschoten}}]{Drögeler2016}%
  \BibitemOpen
  \bibfield  {author} {\bibinfo {author} {\bibfnamefont {M.}~\bibnamefont {Dr{\"o}geler}}, \bibinfo {author} {\bibfnamefont {C.}~\bibnamefont {Franzen}}, \bibinfo {author} {\bibfnamefont {F.}~\bibnamefont {Volmer}}, \bibinfo {author} {\bibfnamefont {T.}~\bibnamefont {Pohlmann}}, \bibinfo {author} {\bibfnamefont {L.}~\bibnamefont {Banszerus}}, \bibinfo {author} {\bibfnamefont {M.}~\bibnamefont {Wolter}}, \bibinfo {author} {\bibfnamefont {K.}~\bibnamefont {Watanabe}}, \bibinfo {author} {\bibfnamefont {T.}~\bibnamefont {Taniguchi}}, \bibinfo {author} {\bibfnamefont {C.}~\bibnamefont {Stampfer}},\ and\ \bibinfo {author} {\bibfnamefont {B.}~\bibnamefont {Beschoten}},\ }\bibfield  {title} {\bibinfo {title} {Spin lifetimes exceeding 12 ns in graphene nonlocal spin valve devices},\ }\href {https://doi.org/10.1021/acs.nanolett.6b00497} {\bibfield  {journal} {\bibinfo  {journal} {Nano Lett.}\ }\textbf {\bibinfo {volume} {16}},\ \bibinfo {pages} {3533} (\bibinfo {year} {2016})}\BibitemShut {NoStop}%
\bibitem [{\citenamefont {Thomsen}\ \emph {et~al.}(2015)\citenamefont {Thomsen}, \citenamefont {Ervasti}, \citenamefont {Harju},\ and\ \citenamefont {Pedersen}}]{PhysRevB.92.195408}%
  \BibitemOpen
  \bibfield  {author} {\bibinfo {author} {\bibfnamefont {M.~R.}\ \bibnamefont {Thomsen}}, \bibinfo {author} {\bibfnamefont {M.~M.}\ \bibnamefont {Ervasti}}, \bibinfo {author} {\bibfnamefont {A.}~\bibnamefont {Harju}},\ and\ \bibinfo {author} {\bibfnamefont {T.~G.}\ \bibnamefont {Pedersen}},\ }\bibfield  {title} {\bibinfo {title} {Spin relaxation in hydrogenated graphene},\ }\href {https://doi.org/10.1103/PhysRevB.92.195408} {\bibfield  {journal} {\bibinfo  {journal} {Phys. Rev. B}\ }\textbf {\bibinfo {volume} {92}},\ \bibinfo {pages} {195408} (\bibinfo {year} {2015})}\BibitemShut {NoStop}%
\bibitem [{\citenamefont {Tuan}\ \emph {et~al.}(2014)\citenamefont {Tuan}, \citenamefont {Ortmann}, \citenamefont {Soriano}, \citenamefont {Valenzuela},\ and\ \citenamefont {Roche}}]{Tuan2014}%
  \BibitemOpen
  \bibfield  {author} {\bibinfo {author} {\bibfnamefont {D.~V.}\ \bibnamefont {Tuan}}, \bibinfo {author} {\bibfnamefont {F.}~\bibnamefont {Ortmann}}, \bibinfo {author} {\bibfnamefont {D.}~\bibnamefont {Soriano}}, \bibinfo {author} {\bibfnamefont {S.~O.}\ \bibnamefont {Valenzuela}},\ and\ \bibinfo {author} {\bibfnamefont {S.}~\bibnamefont {Roche}},\ }\bibfield  {title} {\bibinfo {title} {Pseudospin-driven spin relaxation mechanism in graphene},\ }\href {https://doi.org/10.1038/nphys3083} {\bibfield  {journal} {\bibinfo  {journal} {Nat. Phys.}\ }\textbf {\bibinfo {volume} {10}},\ \bibinfo {pages} {857} (\bibinfo {year} {2014})}\BibitemShut {NoStop}%
\bibitem [{\citenamefont {Van~Tuan}\ \emph {et~al.}(2016)\citenamefont {Van~Tuan}, \citenamefont {Ortmann}, \citenamefont {Cummings}, \citenamefont {Soriano},\ and\ \citenamefont {Roche}}]{VanTuan2016}%
  \BibitemOpen
  \bibfield  {author} {\bibinfo {author} {\bibfnamefont {D.}~\bibnamefont {Van~Tuan}}, \bibinfo {author} {\bibfnamefont {F.}~\bibnamefont {Ortmann}}, \bibinfo {author} {\bibfnamefont {A.~W.}\ \bibnamefont {Cummings}}, \bibinfo {author} {\bibfnamefont {D.}~\bibnamefont {Soriano}},\ and\ \bibinfo {author} {\bibfnamefont {S.}~\bibnamefont {Roche}},\ }\bibfield  {title} {\bibinfo {title} {Spin dynamics and relaxation in graphene dictated by electron-hole puddles},\ }\href {https://doi.org/10.1038/srep21046} {\bibfield  {journal} {\bibinfo  {journal} {Sci. Rep.}\ }\textbf {\bibinfo {volume} {6}},\ \bibinfo {pages} {21046} (\bibinfo {year} {2016})}\BibitemShut {NoStop}%
\bibitem [{\citenamefont {Cummings}\ and\ \citenamefont {Roche}(2016)}]{PhysRevLett.116.086602}%
  \BibitemOpen
  \bibfield  {author} {\bibinfo {author} {\bibfnamefont {A.~W.}\ \bibnamefont {Cummings}}\ and\ \bibinfo {author} {\bibfnamefont {S.}~\bibnamefont {Roche}},\ }\bibfield  {title} {\bibinfo {title} {Effects of dephasing on spin lifetime in ballistic spin-orbit materials},\ }\href {https://doi.org/10.1103/PhysRevLett.116.086602} {\bibfield  {journal} {\bibinfo  {journal} {Phys. Rev. Lett.}\ }\textbf {\bibinfo {volume} {116}},\ \bibinfo {pages} {086602} (\bibinfo {year} {2016})}\BibitemShut {NoStop}%
\bibitem [{\citenamefont {Gebeyehu}\ \emph {et~al.}(2019)\citenamefont {Gebeyehu}, \citenamefont {Parui}, \citenamefont {Sierra}, \citenamefont {Timmermans}, \citenamefont {Esplandiu}, \citenamefont {Brems}, \citenamefont {Huyghebaert}, \citenamefont {Garello}, \citenamefont {Costache},\ and\ \citenamefont {Valenzuela}}]{Gebeyehu_2019}%
  \BibitemOpen
  \bibfield  {author} {\bibinfo {author} {\bibfnamefont {Z.~M.}\ \bibnamefont {Gebeyehu}}, \bibinfo {author} {\bibfnamefont {S.}~\bibnamefont {Parui}}, \bibinfo {author} {\bibfnamefont {J.~F.}\ \bibnamefont {Sierra}}, \bibinfo {author} {\bibfnamefont {M.}~\bibnamefont {Timmermans}}, \bibinfo {author} {\bibfnamefont {M.~J.}\ \bibnamefont {Esplandiu}}, \bibinfo {author} {\bibfnamefont {S.}~\bibnamefont {Brems}}, \bibinfo {author} {\bibfnamefont {C.}~\bibnamefont {Huyghebaert}}, \bibinfo {author} {\bibfnamefont {K.}~\bibnamefont {Garello}}, \bibinfo {author} {\bibfnamefont {M.~V.}\ \bibnamefont {Costache}},\ and\ \bibinfo {author} {\bibfnamefont {S.~O.}\ \bibnamefont {Valenzuela}},\ }\bibfield  {title} {\bibinfo {title} {Spin communication over 30 µm long channels of chemical vapor deposited graphene on sio2},\ }\href {https://doi.org/10.1088/2053-1583/ab1874} {\bibfield  {journal} {\bibinfo  {journal} {2D Mater.}\ }\textbf {\bibinfo {volume} {6}},\ \bibinfo {pages} {034003} (\bibinfo {year}
  {2019})}\BibitemShut {NoStop}%
\bibitem [{\citenamefont {Sousa}\ \emph {et~al.}(2022)\citenamefont {Sousa}, \citenamefont {Perkins},\ and\ \citenamefont {Ferreira}}]{Sousa2022}%
  \BibitemOpen
  \bibfield  {author} {\bibinfo {author} {\bibfnamefont {F.}~\bibnamefont {Sousa}}, \bibinfo {author} {\bibfnamefont {D.~T.~S.}\ \bibnamefont {Perkins}},\ and\ \bibinfo {author} {\bibfnamefont {A.}~\bibnamefont {Ferreira}},\ }\bibfield  {title} {\bibinfo {title} {Weak localisation driven by pseudospin-spin entanglement},\ }\href {https://doi.org/10.1038/s42005-022-01066-z} {\bibfield  {journal} {\bibinfo  {journal} {Commun. Phys.}\ }\textbf {\bibinfo {volume} {5}},\ \bibinfo {pages} {291} (\bibinfo {year} {2022})}\BibitemShut {NoStop}%
\bibitem [{\citenamefont {Dueñas}\ \emph {et~al.}(2023)\citenamefont {Dueñas}, \citenamefont {García},\ and\ \citenamefont {Roche}}]{duenas2023emerging}%
  \BibitemOpen
  \bibfield  {author} {\bibinfo {author} {\bibfnamefont {J.~M.}\ \bibnamefont {Dueñas}}, \bibinfo {author} {\bibfnamefont {J.~H.}\ \bibnamefont {García}},\ and\ \bibinfo {author} {\bibfnamefont {S.}~\bibnamefont {Roche}},\ }\href@noop {} {\bibinfo {title} {{Emerging Spin-Orbit Torques in Low Dimensional Dirac Materials}}} (\bibinfo {year} {2023}),\ \Eprint {https://arxiv.org/abs/2310.06447} {arXiv:2310.06447 [cond-mat.mes-hall]} \BibitemShut {NoStop}%
\bibitem [{\citenamefont {de~Moraes}\ \emph {et~al.}(2020)\citenamefont {de~Moraes}, \citenamefont {Cummings},\ and\ \citenamefont {Roche}}]{PhysRevB.102.041403}%
  \BibitemOpen
  \bibfield  {author} {\bibinfo {author} {\bibfnamefont {B.~G.}\ \bibnamefont {de~Moraes}}, \bibinfo {author} {\bibfnamefont {A.~W.}\ \bibnamefont {Cummings}},\ and\ \bibinfo {author} {\bibfnamefont {S.}~\bibnamefont {Roche}},\ }\bibfield  {title} {\bibinfo {title} {Emergence of intraparticle entanglement and time-varying violation of bell's inequality in dirac matter},\ }\href {https://doi.org/10.1103/PhysRevB.102.041403} {\bibfield  {journal} {\bibinfo  {journal} {Phys. Rev. B}\ }\textbf {\bibinfo {volume} {102}},\ \bibinfo {pages} {041403} (\bibinfo {year} {2020})}\BibitemShut {NoStop}%
\bibitem [{\citenamefont {Yönaç}\ \emph {et~al.}(2007)\citenamefont {Yönaç}, \citenamefont {Yu},\ and\ \citenamefont {Eberly}}]{Yonac2007}%
  \BibitemOpen
  \bibfield  {author} {\bibinfo {author} {\bibfnamefont {M.}~\bibnamefont {Yönaç}}, \bibinfo {author} {\bibfnamefont {T.}~\bibnamefont {Yu}},\ and\ \bibinfo {author} {\bibfnamefont {J.~H.}\ \bibnamefont {Eberly}},\ }\bibfield  {title} {\bibinfo {title} {{Pairwise concurrence dynamics: a four-qubit model}},\ }\href {https://doi.org/10.1088/0953-4075/40/9/S02} {\bibfield  {journal} {\bibinfo  {journal} {J. Phys. B}\ }\textbf {\bibinfo {volume} {40}},\ \bibinfo {pages} {S45} (\bibinfo {year} {2007})}\BibitemShut {NoStop}%
\bibitem [{\citenamefont {Adhikari}\ \emph {et~al.}(2010)\citenamefont {Adhikari}, \citenamefont {Majumdar}, \citenamefont {Home},\ and\ \citenamefont {Pan}}]{swapintrainter}%
  \BibitemOpen
  \bibfield  {author} {\bibinfo {author} {\bibfnamefont {S.}~\bibnamefont {Adhikari}}, \bibinfo {author} {\bibfnamefont {A.~S.}\ \bibnamefont {Majumdar}}, \bibinfo {author} {\bibfnamefont {D.}~\bibnamefont {Home}},\ and\ \bibinfo {author} {\bibfnamefont {A.~K.}\ \bibnamefont {Pan}},\ }\bibfield  {title} {\bibinfo {title} {{Swapping path-spin intraparticle entanglement onto spin-spin interparticle entanglement}},\ }\href {https://doi.org/10.1209/0295-5075/89/10005} {\bibfield  {journal} {\bibinfo  {journal} {Europhys. Lett.}\ }\textbf {\bibinfo {volume} {89}},\ \bibinfo {pages} {10005} (\bibinfo {year} {2010})}\BibitemShut {NoStop}%
\bibitem [{\citenamefont {Kindermann}(2009)}]{PhysRevB.79.115444}%
  \BibitemOpen
  \bibfield  {author} {\bibinfo {author} {\bibfnamefont {M.}~\bibnamefont {Kindermann}},\ }\bibfield  {title} {\bibinfo {title} {Pseudospin entanglement and bell test in graphene},\ }\href {https://doi.org/10.1103/PhysRevB.79.115444} {\bibfield  {journal} {\bibinfo  {journal} {Phys. Rev. B}\ }\textbf {\bibinfo {volume} {79}},\ \bibinfo {pages} {115444} (\bibinfo {year} {2009})}\BibitemShut {NoStop}%
\bibitem [{\citenamefont {Bittencourt}\ and\ \citenamefont {Bernardini}(2017)}]{PhysRevB.95.195145}%
  \BibitemOpen
  \bibfield  {author} {\bibinfo {author} {\bibfnamefont {V.~A. S.~V.}\ \bibnamefont {Bittencourt}}\ and\ \bibinfo {author} {\bibfnamefont {A.~E.}\ \bibnamefont {Bernardini}},\ }\bibfield  {title} {\bibinfo {title} {Lattice-layer entanglement in bernal-stacked bilayer graphene},\ }\href {https://doi.org/10.1103/PhysRevB.95.195145} {\bibfield  {journal} {\bibinfo  {journal} {Phys. Rev. B}\ }\textbf {\bibinfo {volume} {95}},\ \bibinfo {pages} {195145} (\bibinfo {year} {2017})}\BibitemShut {NoStop}%
\bibitem [{\citenamefont {Rashba}(2009)}]{PhysRevB.79.161409}%
  \BibitemOpen
  \bibfield  {author} {\bibinfo {author} {\bibfnamefont {E.~I.}\ \bibnamefont {Rashba}},\ }\bibfield  {title} {\bibinfo {title} {Graphene with structure-induced spin-orbit coupling: Spin-polarized states, spin zero modes, and quantum hall effect},\ }\href {https://doi.org/10.1103/PhysRevB.79.161409} {\bibfield  {journal} {\bibinfo  {journal} {Phys. Rev. B}\ }\textbf {\bibinfo {volume} {79}},\ \bibinfo {pages} {161409} (\bibinfo {year} {2009})}\BibitemShut {NoStop}%
\bibitem [{\citenamefont {Sierra}\ \emph {et~al.}(2021)\citenamefont {Sierra}, \citenamefont {Fabian}, \citenamefont {Kawakami}, \citenamefont {Roche},\ and\ \citenamefont {Valenzuela}}]{Sierra2021}%
  \BibitemOpen
  \bibfield  {author} {\bibinfo {author} {\bibfnamefont {J.~F.}\ \bibnamefont {Sierra}}, \bibinfo {author} {\bibfnamefont {J.}~\bibnamefont {Fabian}}, \bibinfo {author} {\bibfnamefont {R.~K.}\ \bibnamefont {Kawakami}}, \bibinfo {author} {\bibfnamefont {S.}~\bibnamefont {Roche}},\ and\ \bibinfo {author} {\bibfnamefont {S.~O.}\ \bibnamefont {Valenzuela}},\ }\bibfield  {title} {\bibinfo {title} {Van der waals heterostructures for spintronics and opto-spintronics},\ }\href {https://doi.org/10.1038/s41565-021-00936-x} {\bibfield  {journal} {\bibinfo  {journal} {Nat. Nanotechnol.}\ }\textbf {\bibinfo {volume} {16}},\ \bibinfo {pages} {856} (\bibinfo {year} {2021})}\BibitemShut {NoStop}%
\bibitem [{\citenamefont {Castro~Neto}\ \emph {et~al.}(2009)\citenamefont {Castro~Neto}, \citenamefont {Guinea}, \citenamefont {Peres}, \citenamefont {Novoselov},\ and\ \citenamefont {Geim}}]{CastroNeto2009}%
  \BibitemOpen
  \bibfield  {author} {\bibinfo {author} {\bibfnamefont {A.~H.}\ \bibnamefont {Castro~Neto}}, \bibinfo {author} {\bibfnamefont {F.}~\bibnamefont {Guinea}}, \bibinfo {author} {\bibfnamefont {N.~M.~R.}\ \bibnamefont {Peres}}, \bibinfo {author} {\bibfnamefont {K.~S.}\ \bibnamefont {Novoselov}},\ and\ \bibinfo {author} {\bibfnamefont {A.~K.}\ \bibnamefont {Geim}},\ }\bibfield  {title} {\bibinfo {title} {The electronic properties of graphene},\ }\href {https://doi.org/10.1103/RevModPhys.81.109} {\bibfield  {journal} {\bibinfo  {journal} {Rev. Mod. Phys.}\ }\textbf {\bibinfo {volume} {81}},\ \bibinfo {pages} {109} (\bibinfo {year} {2009})}\BibitemShut {NoStop}%
\bibitem [{\citenamefont {Zarea}\ and\ \citenamefont {Sandler}(2009)}]{Zarea2009}%
  \BibitemOpen
  \bibfield  {author} {\bibinfo {author} {\bibfnamefont {M.}~\bibnamefont {Zarea}}\ and\ \bibinfo {author} {\bibfnamefont {N.}~\bibnamefont {Sandler}},\ }\bibfield  {title} {\bibinfo {title} {Rashba spin-orbit interaction in graphene and zigzag nanoribbons},\ }\href {https://doi.org/10.1103/PhysRevB.79.165442} {\bibfield  {journal} {\bibinfo  {journal} {Phys. Rev. B}\ }\textbf {\bibinfo {volume} {79}},\ \bibinfo {pages} {165442} (\bibinfo {year} {2009})}\BibitemShut {NoStop}%
\bibitem [{\citenamefont {Wootters}(1998)}]{Wootters1998}%
  \BibitemOpen
  \bibfield  {author} {\bibinfo {author} {\bibfnamefont {W.~K.}\ \bibnamefont {Wootters}},\ }\bibfield  {title} {\bibinfo {title} {Entanglement of formation of an arbitrary state of two qubits},\ }\href {https://doi.org/10.1103/PhysRevLett.80.2245} {\bibfield  {journal} {\bibinfo  {journal} {Phys. Rev. Lett.}\ }\textbf {\bibinfo {volume} {80}},\ \bibinfo {pages} {2245} (\bibinfo {year} {1998})}\BibitemShut {NoStop}%
\bibitem [{\citenamefont {Jacoboni}\ and\ \citenamefont {Reggiani}(1983)}]{Jacoboni1983}%
  \BibitemOpen
  \bibfield  {author} {\bibinfo {author} {\bibfnamefont {C.}~\bibnamefont {Jacoboni}}\ and\ \bibinfo {author} {\bibfnamefont {L.}~\bibnamefont {Reggiani}},\ }\bibfield  {title} {\bibinfo {title} {The monte carlo method for the solution of charge transport in semiconductors with applications to covalent materials},\ }\href {https://doi.org/10.1103/RevModPhys.55.645} {\bibfield  {journal} {\bibinfo  {journal} {Rev. Mod. Phys.}\ }\textbf {\bibinfo {volume} {55}},\ \bibinfo {pages} {645} (\bibinfo {year} {1983})}\BibitemShut {NoStop}%
\bibitem [{sup()}]{suppmaterial}%
  \BibitemOpen
  \href@noop {} {}\bibinfo {note} {See Supplemental Material at [URL will be inserted by publisher] for a detailed description of the Monte Carlo simulations, an analytical derivation of the EY spin relaxation time, and a derivation of the saturated concurrence bounds.}\BibitemShut {Stop}%
\bibitem [{\citenamefont {Reimann}(2008)}]{Reimann2008}%
  \BibitemOpen
  \bibfield  {author} {\bibinfo {author} {\bibfnamefont {P.}~\bibnamefont {Reimann}},\ }\bibfield  {title} {\bibinfo {title} {Foundation of statistical mechanics under experimentally realistic conditions},\ }\href {https://doi.org/10.1103/PhysRevLett.101.190403} {\bibfield  {journal} {\bibinfo  {journal} {Phys. Rev. Lett.}\ }\textbf {\bibinfo {volume} {101}},\ \bibinfo {pages} {190403} (\bibinfo {year} {2008})}\BibitemShut {NoStop}%
\bibitem [{\citenamefont {Short}(2011)}]{Short2011}%
  \BibitemOpen
  \bibfield  {author} {\bibinfo {author} {\bibfnamefont {A.~J.}\ \bibnamefont {Short}},\ }\bibfield  {title} {\bibinfo {title} {Equilibration of quantum systems and subsystems},\ }\href {https://doi.org/10.1088/1367-2630/13/5/053009} {\bibfield  {journal} {\bibinfo  {journal} {New J. Phys.}\ }\textbf {\bibinfo {volume} {13}},\ \bibinfo {pages} {053009} (\bibinfo {year} {2011})}\BibitemShut {NoStop}%
\bibitem [{\citenamefont {Richter}\ \emph {et~al.}(2019)\citenamefont {Richter}, \citenamefont {Lamann}, \citenamefont {Bartsch}, \citenamefont {Steinigeweg},\ and\ \citenamefont {Gemmer}}]{Richter2019}%
  \BibitemOpen
  \bibfield  {author} {\bibinfo {author} {\bibfnamefont {J.}~\bibnamefont {Richter}}, \bibinfo {author} {\bibfnamefont {M.~H.}\ \bibnamefont {Lamann}}, \bibinfo {author} {\bibfnamefont {C.}~\bibnamefont {Bartsch}}, \bibinfo {author} {\bibfnamefont {R.}~\bibnamefont {Steinigeweg}},\ and\ \bibinfo {author} {\bibfnamefont {J.}~\bibnamefont {Gemmer}},\ }\bibfield  {title} {\bibinfo {title} {Relaxation of dynamically prepared out-of-equilibrium initial states within and beyond linear response theory},\ }\href {https://doi.org/10.1103/PhysRevE.100.032124} {\bibfield  {journal} {\bibinfo  {journal} {Phys. Rev. E}\ }\textbf {\bibinfo {volume} {100}},\ \bibinfo {pages} {032124} (\bibinfo {year} {2019})}\BibitemShut {NoStop}%
\bibitem [{\citenamefont {\ifmmode \check{Z}\else \v{Z}\fi{}uti\ifmmode~\acute{c}\else \'{c}\fi{}}\ \emph {et~al.}(2004)\citenamefont {\ifmmode \check{Z}\else \v{Z}\fi{}uti\ifmmode~\acute{c}\else \'{c}\fi{}}, \citenamefont {Fabian},\ and\ \citenamefont {Das~Sarma}}]{Zutic2004}%
  \BibitemOpen
  \bibfield  {author} {\bibinfo {author} {\bibfnamefont {I.}~\bibnamefont {\ifmmode \check{Z}\else \v{Z}\fi{}uti\ifmmode~\acute{c}\else \'{c}\fi{}}}, \bibinfo {author} {\bibfnamefont {J.}~\bibnamefont {Fabian}},\ and\ \bibinfo {author} {\bibfnamefont {S.}~\bibnamefont {Das~Sarma}},\ }\bibfield  {title} {\bibinfo {title} {Spintronics: Fundamentals and applications},\ }\href {https://doi.org/10.1103/RevModPhys.76.323} {\bibfield  {journal} {\bibinfo  {journal} {Rev. Mod. Phys.}\ }\textbf {\bibinfo {volume} {76}},\ \bibinfo {pages} {323} (\bibinfo {year} {2004})}\BibitemShut {NoStop}%
\bibitem [{\citenamefont {Nguyen}\ \emph {et~al.}(2022)\citenamefont {Nguyen}, \citenamefont {Hoang},\ and\ \citenamefont {Charlier}}]{Nguyen_2022}%
  \BibitemOpen
  \bibfield  {author} {\bibinfo {author} {\bibfnamefont {V.~H.}\ \bibnamefont {Nguyen}}, \bibinfo {author} {\bibfnamefont {T.~X.}\ \bibnamefont {Hoang}},\ and\ \bibinfo {author} {\bibfnamefont {J.-C.}\ \bibnamefont {Charlier}},\ }\bibfield  {title} {\bibinfo {title} {Electronic properties of twisted multilayer graphene},\ }\href {https://doi.org/10.1088/2515-7639/ac6c4a} {\bibfield  {journal} {\bibinfo  {journal} {Journal of Physics: Materials}\ }\textbf {\bibinfo {volume} {5}},\ \bibinfo {pages} {034003} (\bibinfo {year} {2022})}\BibitemShut {NoStop}%
\end{thebibliography}
\end{document}